\documentclass[fleqn,twoside]{article}
\usepackage{espcrc2}

\usepackage{graphicx}

\def\pbp{\rm{pbp}}

\title{Precise Measurement of the Absolute Yield of Fluorescence
  Photons in Atmospheric Gases}

\author{{\bf AIRFLY Collaboration}: M.~Ave\address[IK]{Karlsruhe Institute of Technology, IK, Postfach 6980, D
    - 76021 Karlsruhe, Germany},
  M.~Boh\'a\v{c}ov\'a\address[Chi]{University of
    Chicago, Enrico Fermi Institute \& Kavli Institute for
    Cosmological Physics,\\ 5640 S. Ellis Ave., Chicago, IL 60637,
    USA}\address[CZ1]{Institute of Physics of the Academy of
    Sciences of the Czech Republic,\\ Na Slovance 2, CZ-182 21 Praha 8,
    Czech Republic},
        K. Daumiller\addressmark[IK],
    P.~Di Carlo\address[AqU]{Dipartimento di Fisica dell'Universit\`{a}
    de l'Aquila and INFN, Via Vetoio, I-67010 Coppito, Aquila, Italy}, 
  C.~Di Giulio\address[RomeU]{Dipartimento di Fisica
    dell'Universit\`{a} di Roma Tor Vergata and Sezione INFN,\\ Via
    della Ricerca Scientifica, I-00133 Roma, Italy}, 
   P.~Facal San Luis\addressmark[Chi]\thanks{Corresponding author: {\tt facal@kicp.uchicago.edu}},
D.~Gonzales\address[IEKP]{Karlsruhe Institute of Technology, IEKP, Postfach 3640, D - 76021 Karlsruhe, Germany},
    C.~Hojvat\address[FNAL]{Fermi National Accelerator Laboratory, Batavia, IL 60510, USA},
    J.~R.~H\"orandel\address[IMAPP]{IMAPP, Radboud University Nijmegen, 6500 GL Nijmegen, The Netherlands},
  M.~Hrabovsk\'y\address[CZ2]{Palacky University, RCATM, Olomuc, Czech Republic},
  M.~Iarlori\addressmark[AqU],
    B.~Keilhauer\addressmark[IK],
  H.~Klages\addressmark[IK],
  M.~Kleifges\address[IPE]{Karlsruhe Institute of Technology, IPE, Postfach 3640, D - 76021 Karlsruhe, Germany},
  F.~Kuehn\addressmark[FNAL],
   M.~Monasor\addressmark[Chi],
  L.~No\v{z}ka\addressmark[CZ1],
   M.~Palatka\addressmark[CZ1],
  S.~Petrera\addressmark[AqU],
  P.~Privitera\addressmark[Chi],
  J.~Ridky\addressmark[CZ1],
  V.~Rizi\addressmark[AqU],
   B.~Rouill\'e d'Orfeuil\addressmark[Chi],
   F.~Salamida\addressmark[IK],
  P.~Schov\'anek\addressmark[CZ1],
  R.~\v{S}mida\addressmark[IK],
  H.~Spinka\address[ANL]{Argonne National Laboratory, Argonne, IL
  60439, USA},
   A.~Ulrich\address[Mun]{Physik Department E12, Technische
    Universit\"{a}t Muenchen,\\ James Franck Str. 1, D-85748 Garching,
    Germany},
   V.~Verzi\addressmark[RomeU],
    C.~Williams\addressmark[Chi]
 }

\begin{document}

\begin{abstract}
We have performed a measurement of the absolute yield of fluorescence photons at the Fermilab Test Beam. A systematic uncertainty at 5\% level was achieved  by the use of Cherenkov radiation as a reference calibration light source. A cross-check was performed by an independent calibration using a laser light source. A significant improvement on the energy scale uncertainty of Ultra-High Energy Cosmic Rays  is expected.
\vspace{1pc}
\end{abstract}

% typeset front matter (including abstract)
\maketitle

\section{Introduction}
Fluorescence detection of Ultra High Energy Cosmic Rays (UHECRs) is a well established technique, pioneered by Fly's
Eye~\cite{Baltrusaitis:1985mx}, and today 
an integral part of the Pierre Auger~\cite{Abraham:2009pm} and the
Telescope Array~\cite{Tameda:2009zza} experiments.  Excitation of the
atmospheric nitrogen by the charged particles in the extensive air
shower induces the emission of fluorescence photons, mostly in the
300-400~nm range. A Fluorescence Detector (FD) records this radiation to
infer the cosmic ray energy and the particle type. For this purpose, the
fluorescence light yield from the charged particles in the shower must
be known for every emission point along the shower axis. A correction
is then applied to account for atmospheric effects between the shower
and the telescope, enabling an accurate, \emph{quasi-calorimetric},
primary energy determination.

The uncertainty on the fluorescence light yield is one of the main
systematic uncertainties on the cosmic ray energy determination by
experiments that employ the fluorescence technique (e.g. 14\% over a total 22\% uncertainty for the Pierre Auger experiment). The AIRFLY
collaboration has already performed a very precise measurement of the
fluorescence spectrum and its pressure dependence~\cite{Ave:2007xh},
as well as the dependence of the emission on the temperature and
humidity~\cite{Ave:2007em}. AIRFLY  measurements
over electron kinetic energies ranging from keV to GeV
using several accelerators have also proven the proportionality of the
fluorescence yield with the electron energy deposit~\cite{Ave:2008zz}.

The final step in the precise characterization of the nitrogen
fluorescence light emission is the measurement the absolute value of
the yield for the main emission line at 337~nm. The AIRFLY strategy to reduce the
systematic uncertainties is to calibrate the
experimental apparatus in situ, using photons emitted by a well know
process: Cherenkov radiation~\cite{Bohacova:2008vg}. A second
calibration method, with nearly independent systematic uncertainty, is based an absolutely calibrated  laser light source. In this paper, we present preliminary results from a series of
dedicated measurements at the Fermi National Accelerator Laboratory
(Fermilab).

\section{Experimental method}
\begin{figure}
\includegraphics[width=\columnwidth]{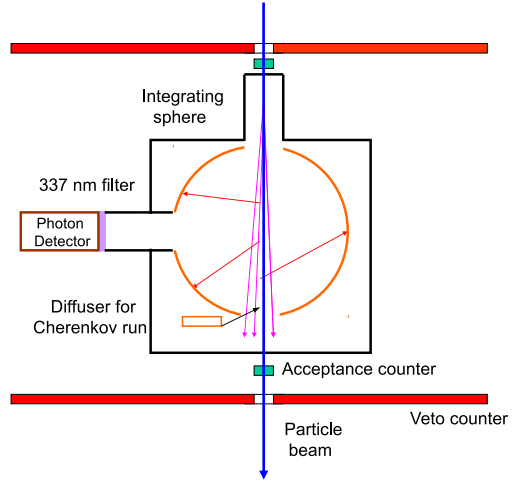}
\caption{Layout of the experimental apparatus used at the Fermilab
  Meson Test Beam\label{layout}}
\end{figure}
The measurements were performed at the Fermilab Test Beam
Facility. Most of the measurements were carried out using the 120 GeV
proton beam, extracted from the Main Injector. Secondary beams of 32
GeV pions and 8 GeV positrons were also used.

\begin{figure}
\includegraphics[width=\columnwidth]{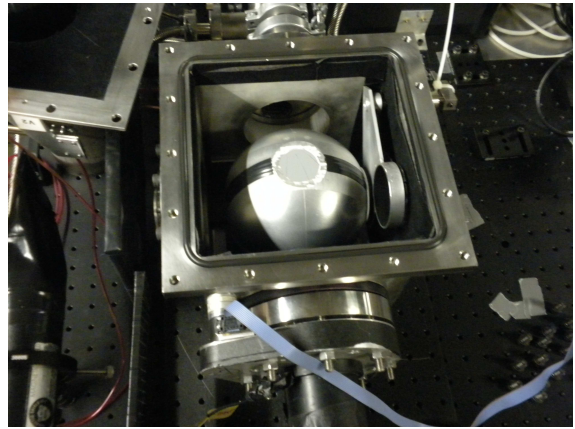}
\caption{AIRFLY chamber with the integrating sphere fitted
  inside.\label{chamber}}
\end{figure}
A sketch of the apparatus is
shown in Fig. \ref{layout}.  A fluorescence chamber made of a 3 mm
thick stainless steel was placed in the beam path, with the
corresponding flanges for windows, shutters, gauges, gas inlet and
pump-out.  Both the entrance and exit windows are 0.1 mm thick aluminum,
with the entrance window before a 18 cm long tube to provide
additional length for Cherenkov light production. An integrating
sphere was used to collect light produced inside the chamber
(Fig.~\ref{chamber}).  One of the ports of the sphere was open to a a
gas-tight window fitted with a 337~nm filter and then coupled to the
photon detector.  A Hamamatsu H7195P photomultiplier (PMT) tube, with
good single photoelectron resolution, was used for photon
detection. The optical field of view was defined by a 40 mm diameter
acceptance cylinder placed between the integrating sphere's port and
the filter, and by circular apertures of the same size placed in front
of the PMT photocathode. A mechanical shutter remotely controlled
allowed to take background measurements. The chamber was internally
covered with a black UV-absorbing material to avoid stray light.

\begin{figure}
  \begin{center}
    \includegraphics[width=.8\columnwidth]{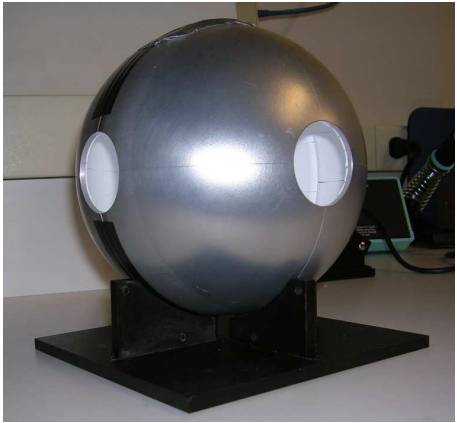}
    \caption{The integrating sphere used in the measurements. The
      diffusive coating applied to the interior can be seen through two of
      the ports.\label{sphere}}
  \end{center}
\end{figure}

The purpose of the integrating sphere is twofold:  it increases the light collection solid angle (and thus
maximizes the signal to noise ratio) and it works as
equalizer of the detection efficiency of the set-up for the isotropic
fluorescence light and for the highly directional Cherenkov light. The
integrating sphere (Fig.~\ref{sphere}) was built from two hollow
aluminum half spheres coated with a material of very high diffusive
reflectance. Light produced inside the sphere is collected over 4$\pi$
solid angle and at the same time isotropized by several diffusion
bounces, so that the lambertian light output is independent of the
original light distribution. A total of 4 ports were machined in the
sphere, one was the detection port, two were for the beam entrance and
exit, and the last one was opened at the top of the sphere.  Two
remotely controlled shutters could close the top and exit ports of the
integrating sphere with plugs coated with the same diffusive material
material using inside the sphere. With the exit port closed the
Cherenkov is diffused back into the sphere and can reach the photon
detector. With the exit port open the Cherenkov is absorbed by the
chamber lining and thus only fluorescence can reach the photon
detector. The top port compensates the open/closed position of the
exit port to maintain always three open ports and thus the optical
characteristics of the sphere. 
%In some configurations used during the
%test beam, and additional UV-absorbing cylinder was placed after the
%open port to further reduce the amount of Cherenkov reflected back
i%nto the sphere when the exit port was open.

The whole chamber was airtight and a remotely controlled system for
gas and vacuum handling was used. Pure nitrogen and a dry-air mixture
were used for the measurements and additionally helium and argon were
used for background runs. The pressure, temperature and humidity
inside the chamber were monitored remotely using the appropriate
sensors.

%\begin{figure}
%\includegraphics[width=\columnwidth]{counters}
%\caption{The finger (front) and one of the veto counters placed at the entrance
%  of the chamber. \label{counters}}
%\end{figure}

A set of particle counters was used for beam monitoring. At the
entrance of the chamber a 10 mm diameter finger counter %(Fig.~\ref{counters})
was used for beam tagging. At the exit the beam was tagged with a
Cherenkov counter, a 10~mm diameter cylindrical rod made of
UV-transparent acrylic material. The rod was 30~mm long and allowed
very good single particle resolution with fast timing. Before and
after the chamber two big scintillator pads with a central 10~mm diameter hole
for beam passage provided a veto for off-axis particles. The chamber
and the counters were mounted on an optical breadboard for precise
mounting and alignment, and placed on a remotely movable table that
allowed the alignment of the apparatus with the beam by maximizing the
finger and Cherenkov counters rates.  The beam profile was monitored
by wire chambers placed before and after the AIRFLY apparatus and was
typically 3 mm x 4 mm wide.

\begin{figure}[!ht]
\includegraphics[width=\columnwidth]{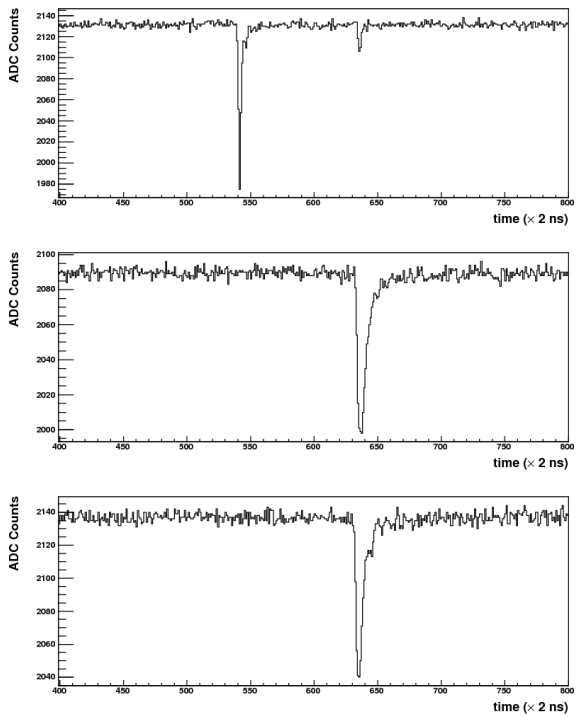}
\caption{The signal of a train of bunches for the Cherenkov rod (top
  plot) and the two veto counters (middle and bottom plots). Two particles
  in two different bunches can be observed: the first one passes
  through the rod while the second one passes through both veto
  counters (and leaves also a small signal in the rod when it hits the
  PMT glass producing a small amount of Cherenkov in
  it).\label{train}}
\end{figure}

The trigger and DAQ were designed considering the characteristics of
the beam timing: within a 4~s beam spill, particles were grouped in
trains of bunches. Trains were separated in time by 10 $\mu s$,
and bunches within a train were separated by 19 ns.  Typical
conditions for data taking with the proton beam were 30 bunches per
train, and a multiplicity of $2\cdot 10^5$ particles per spill. The
trigger logic was built from the coincidence of a train trigger gate,
issued in correspondence of the arrival of each train of bunches, and
a single particle trigger gate, coming from a beam monitoring
scintillator counter. Both triggers were provided by the Test
Beam Facility.  Whenever a trigger was issued the signals from the
scintillator counters and photomultipliers were digitized by a 12-bit
500 MHz FADC and 600 samples (equivalent to 1.2 $\mu$s, containing the
entire train of bunches) were saved in the FADC memory. The data for
the whole spill was stored in the FADC internal memory and then
readout and saved to disk in about 40 s between spills.  As an
example, the ADC trace of the beam counters for one trigger is
shown in Fig. \ref{train}.

Data during the test beam was taken in a different configurations. For
each configuration runs of up to half hour (i.e. 30 spills) of data
were acquired.  Runs in the same conditions were repeated periodically
to improve statistics and to assure redundancy and consistency.

\begin{figure}[!ht]
  \begin{center}
\includegraphics[width=.9\columnwidth]{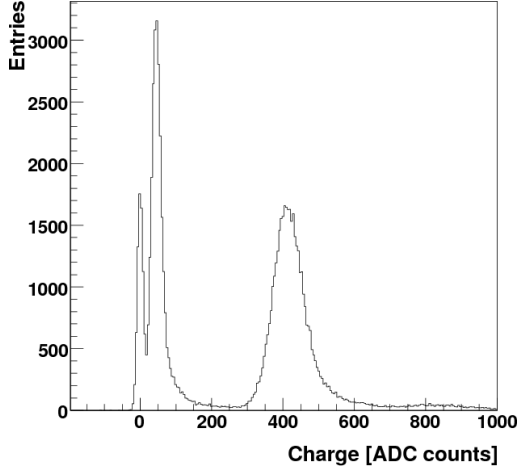}
\caption{Spectrum of the Cherenkov rod. From left to right the three
  peaks correspond to pedestal, particles hitting the PMT window and
  single particles passing through the Cherenkov radiator.  The fourth
  rightmost smaller peak corresponds to two particles in the same
  bunch.
\label{rod}}
  \end{center}
\end{figure}

\begin{figure}[!ht]
  \begin{center}
\includegraphics[width=.9\columnwidth]{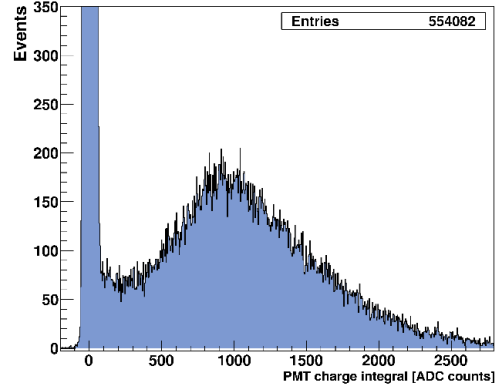}
\caption{PMT spectrum for a run. The events inside the single
  photoelectron peak are counted to define the signal.
\label{photoelec}}
  \end{center}
\end{figure}

\section{Data analysis and results}
 Offline data analysis for one run proceeds selecting single particles
 that cross the fiducial volume of the camera, hence are tagged by the
 beam counters placed at the entrance and the exit of the chamber
 (Fig.~\ref{rod}). For these selected particles we require that no
 signal is present in any of the veto counters. Additional analysis
 cuts are placed to discard trains that have unusually large number of
 particles passing through the veto counters, as this has been shown
 to improve the background conditions without excessively penalizing
 statistics.  Once the clean single protons have been selected the PMT
 signal is analyzed and the photons in coincidence with the selected
 protons are counted (Fig~\ref{photoelec}). The signal $S$, in units
 of photoelectrons per beam particle, \pbp, is then calculated.

The measured signal in the a fluorescence measurement taken in a given
gas, S$_{FL}^{gas}(meas)$, is given by:
 \begin{equation}
S_{FL}^{gas}(meas) = S_{FL}^{gas}+ B_{FL}^{gas},
\label{eq:fluor1}
\end{equation}
where S$^{gas}_{FL}$ is the signal from the 337 nm band, and
B$^{gas}_{FL}$ is attributable to background.  The best determination
of the overall background can be obtained combining measurements from
different gases.  From Eq. \ref{eq:fluor1}, the difference of the
measured signal in $\rm{N_2}$ and air is given by:
 \begin{equation}
 \Delta S_{FL} = S^{\rm{N_2}}_{FL} - S^{\rm{air}}_{FL} +
 B^{\rm{N_2}}_{FL} - B^{\rm{air}}_{FL}.
\label{eq:fluor2}
\end{equation}
Since the beam related background and the primary interactions are the same
and secondary particle production is very similar in nitrogen and air
backgrounds cancel in Eq. \ref{eq:fluor2}. Thus,
 \begin{equation}
 \Delta S_{FL} =  S^{\rm{N_2}}_{FL} \left( 1 - \frac{1}{r_{\rm{N_2}}} \right),
\label{eq:fluor3}
\end{equation}
where $r_{\rm{N_2}}$ is the ratio of the 337 nm fluorescence in pure
nitrogen to the signal in air. This ratio was measured by AIRFLY as a
function of pressure in~\cite{Ave:2007xh} and was cross-checked in our Fermilab 
experimental apparatus using a $^{241}$Am radioactive source, giving
$7.45 \pm 0.10$ at 1000 hPa.

We measure $\Delta S_{FL}=(16.83 \pm 0.13)\cdot
10^{-4}~\pbp$. From Eq.~\ref{eq:fluor3}, we derive the
background-subtracted fluorescence signal:
\begin{equation}
 S^{\rm{N_2}}_{FL} = (19.44 \pm 0.15)\cdot 10^{-4} ~\pbp,
\label{eq:fluorsignn2}
\end{equation}
with the background (Eq.~\ref{eq:fluor1})  estimated as a 3\% of the signal.

%and now from  an estimate of the background 
%\begin{equation}
% B^{\rm{N_2}}_{FL} = (0.61 \pm 0.08)\cdot 10^{-4} ~\pbp.
%\label{eq:fluorbkg}
%\end{equation}
%which is at the level of 3\% of the signal.

In the Cherenkov calibration measurement, both Cherenkov and
fluorescence emission in the gas contribute to the measured signal:
 \begin{equation}
S_{CH}^{gas}(meas) = S^{gas}_{FL}+S^{gas}_{CH}+ B^{gas}_{FL}+B_{CH},
\label{eq:cher1}
\end{equation}
where $S^{gas}_{CH}$ is the signal from Cherenkov light in the 337 nm
band emitted in the gas under study, and $B_{CH}$ takes into account
background originating from the interaction of the beam particles in
the exit port plug (that is independent of the gas filling the
chamber). $B_{CH}$ can be estimated directly from the difference in
the vacuum measurement in fluorescence and Cherenkov modes
\begin{equation}
B_{CH} =(2.63 \pm 0.23) \cdot 10^{-4}~\pbp, 
\label{eq:cherbkg1}
\end{equation}
a value that is $\sim 10$\% of the Cherenkov signal. We made several
cross-check of this  background, for example changing the
exit port plug material to a thin Mylar foil where no light production
is expected.

We measure $S_{CH}^{\rm{N_2}}(meas) = (32.89 \pm 0.15)\cdot
10^{-4}~\pbp$. Using Eq.~\ref{eq:cher1} we obtain 
\begin{equation}
 S^{\rm{N_2}}_{CH} = (10.27 \pm 0.23)\cdot 10^{-4} ~\pbp,
\label{eq:chersign}
\end{equation}
and from it and Eq.~\ref{eq:fluorsignn2} we derive
\begin{equation}
R^{\rm{N_2}} = \frac{S^{\rm{N_2}}_{FL}}{S^{\rm{N_2}}_{CH}} = 1.893 \pm 0.045,
\label{eq:ratio}
\end{equation}
the ratio of fluorescence to Cherenkov 337~nm photons produced inside
the chamber.

In order to derive the absolute 337~nm yield in air, we  performed
a full Monte Carlo simulation of the setup, where all individual components are
simulated according to measurements done in the laboratory.
The absolute fluorescence yield in the simulation, $Y^{\rm{air}}_{MC}$, which determines  a corresponding expected
Fluorescence/Cherenkov ratio $R^{air}_{MC}$, can be scaled to match the measured value of  Eq.~\ref{eq:ratio} and obtain our measurement of the absolute
fluorescence yield in air:
%\begin{equation}
%  Y^{\rm{air}} = \frac{R^{\rm{N_2}}/r_{N_2}}{R^{\rm{air}}_{MC}} Y^{\rm{air}}_{MC} 
%\end{equation}
\begin{equation}
    Y^{\rm{air}}  = 5.60 \pm 0.13 \,\, \gamma_{\rm 337 nm}/{\rm MeV},
\end{equation}
where the uncertainty is statical only. Our preliminary estimate for
the systematic uncertainty of the measurement is better than 5\%,
dominated by the uncertainty in the wavelength dependence of the PMT
quantum efficiency and the 337 filter transmission, accounting for the
different spectral distributions of the fluorescence and Cherenkov
signals. The detailed analysis of the uncertainties is still,
however, ongoing. 

In the laser calibration method, we use a 337~nm nitrogen laser and an
NIST calibrated probe to determine the overall efficiency of the
set-up, the number of detected photoelectrons per photon entering the
sphere. Using this efficiency, the number in Eq.~\ref{eq:fluorsignn2} and
a detailed Monte Carlo simulation of the laser calibration the value
\begin{equation}
    Y^{\rm{air}}_{\rm LASER}  = 5.56 \pm 0.07 \,\, \gamma_{\rm 337 nm}/{\rm MeV},
\end{equation}
is obtained, which has an uncertainty nearly independent of the one from
the Cherenkov calibration. The systematic uncertainty is dominated by
the 5\% uncertainty in the absolute calibration of the laser probe.

\section{Outlook}
The  preliminary results on the absolute fluorescence yield  reported in this work are compatible with the
current values used in Fluorescence Detector analysis (see Ref.~\cite{Rosado:2010au} for a review of the different
measurements.) We expect a final measurement with a systematic
uncertainty below 5\%, a significant improvement over previous measurements,  which will correspondingly  improve the uncertainty on the energy scale of UHECR measurements.

\end{document}